\documentclass[a4paper,aps,prd,10pt,preprintnumbers,showpacs,twocolumn,superscriptaddress,nofootinbib,amsmath,amssymb]{revtex4-1}
\usepackage[dvips]{graphics}
\usepackage{cmap}
\usepackage[utf8]{inputenc}
\usepackage[T1]{fontenc}

\def\imo{i}
\def\Order#1{{\cal O}\left(#1\right)}

\begin{document}
\title{Non-Schwarzschild black-hole metric in four dimensional higher derivative gravity: analytical approximation}

\author{K. Kokkotas}
\email{kostas.kokkotas@uni-tuebingen.de}
\affiliation{Theoretical Astrophysics, Eberhard-Karls University of Tübingen, Tübingen 72076, Germany}
\author{R. A. Konoplya}
\email{roman.konoplya@uni-tuebingen.de}
\affiliation{Theoretical Astrophysics, Eberhard-Karls University of Tübingen, Tübingen 72076, Germany}
\affiliation{Institute  of  Physics  and  Research  Centre  of  Theoretical  Physics  and  Astrophysics, Faculty  of  Philosophy  and  Science,  Silesian  University  in  Opava,  Opava,  Czech  Republic}
\author{A. Zhidenko}
\email{olexandr.zhydenko@ufabc.edu.br}
\affiliation{Centro de Matemática, Computação e Cognição (CMCC), Universidade Federal do ABC (UFABC),\\ Rua Abolição, CEP: 09210-180, Santo André, SP, Brazil}

\begin{abstract}
Higher derivative extensions of Einstein gravity are important within the string theory approach to gravity and as alternative and effective theories of gravity. H.~Lü,  A.~Perkins, C.~Pope, K.~Stelle  [Phys.~Rev.~Lett. \textbf{114} (2015), 171601] found a numerical solution describing a spherically symmetric non-Schwarzschild asymptotically flat black hole in the Einstein gravity with added higher derivative terms. Using the general and quickly convergent parametrization in terms of the continued fractions, we represent this numerical solution in the analytical form, which is accurate not only near the event horizon or far from black hole, but \emph{in the whole space}. Thereby, the obtained analytical form of the metric allows one to study easily all the further properties of the black hole, such as thermodynamics, Hawking radiation, particle motion, accretion, perturbations, stability, quasinormal spectrum, etc. Thus, the found analytical approximate representation can serve in the same way as an exact solution.
\end{abstract}


\maketitle

\section{Introduction}

Recent observation of gravitational waves from, apparently, the binary black holes merger \cite{Abbott:2016blz} and considerable progress in observations of the galactic black hole in the electromagnetic spectrum \cite{Goddi:2016jrs} made black holes important objects for testing the regime of strong gravity.
At the same time, the current lack of accuracy in determination the angular momentum and mass of the resultant ringing black hole leaves open the window for alternative theories of gravity, allowing for deviations from Schwarzschild and Kerr geometries \cite{Konoplya:2016pmh}. One of such interesting alternatives is the Einstein gravity with added quadratic in curvature term for which the most general action has the form
\begin{equation}
\label{HDGaction}
I = \int d^4x\sqrt{-g}\left(\gamma R -
\alpha C_{\mu\nu\rho\sigma}C^{\mu\nu\rho\sigma} + \beta R^{2}\right)\,,
\end{equation}
where $\alpha$, $\beta$ and $\gamma$ are constants, $C_{\mu\nu\rho\sigma}$ is the Weyl tensor.

In \cite{Lu:2015cqa} it was shown that in addition to the Schwarzschild solution in the theory (\ref{HDGaction}), there is another spherically symmetric asymptotically flat non-Schwarzschild black-hole solution within the same theory. The numerical solution for the non-Schwarzschild case was represented in \cite{Lu:2015cqa,Lu:2015psa}.

Numerical solution, although it can be used for further numerical analysis at fixed values of parameters, does not give a clear picture of dependence of the metric on physical parameters of the system. Therefore, the general method of parametrization of black-hole spacetimes was developed in \cite{Rezzolla:2014mua} for spherically symmetric and in \cite{Konoplya:2016jvv} for axially-symmetric black holes.
In the spherically symmetric case, considered here, the method is based on the continued fraction expansion in terms of a compactified radial coordinate. Comparison of observables, such as position of the innermost stable circular orbit and shadows cast by black holes demonstrated that this method turned out to be rapidly convergent \cite{Younsi:2016azx}, giving us an opportunity for finding relatively concise analytical approximation for a black-hole metric.

Here we shall use the above mentioned continued fraction parametrization and find the analytical form for the asymptotically flat non-Schwarzschild numerical solution \cite{Lu:2015cqa} in the Einstein gravity with quadratic in curvature corrections. The obtained metric satisfies the currently existing constrains on the post-Newtonian (weak field) behavior.

The metric functions are represented as a ratio of polynomials of the radial coordinate with the coefficients, which depend on the coupling constant and black-hole radius. The latter re-scales the black hole mass and radial coordinate and can be fixed in further analysis. The main result of our work is the obtained analytical fourth order representation of the metric (written down explicitly in Appendix A), which is accurate in the whole space outside the black hole. This allows one to use it effectively for various studies of the black-hole properties and analysis of interactions between the black hole and surrounding matter.

The paper is organized as follows. Sec.~\ref{sec:theory} briefly discuss the theory under consideration and shows that without loss of generality it can be reduced to the Einstein-Weyl theory when considering spherically symmetric solutions. Sec.~\ref{sec:metric} relates the deduction of the analytical expressions for the metric functions with the help of the continued fraction parametrization. Sec.~\ref{sec:accuracy} is devoted to testing the accuracy of the obtained analytical metric through calculation of observable characteristics: rotational frequency on the innermost stable circular orbit and eikonal quasinormal modes. Finally, in Sec.~\ref{sec:discussion} we discuss the obtained results and spotlight the main potentially interesting applications that can be done based on the obtained here analytical form of the metric.

\section{Static solutions in the Einstein gravity with added quadratic terms}\label{sec:theory}

\subsection{Analytical approximation}

One of the coupling constants can be fixed when choosing the system of units, so we take $\gamma=1$.
Then, the equations of motion take the form
\begin{eqnarray}
R_{\mu\nu}-\frac{1}{2} R g_{\mu\nu} &-& 4 \alpha B_{\mu\nu} +2\beta
 R\left(R_{\mu\nu}-\frac{1}{4} R g_{\mu\nu}\right) \nonumber\\
&+& 2\beta(g_{\mu\nu}\square R-\nabla_\mu\nabla_\nu R)=0\,,\label{eoms}
\end{eqnarray}
where
\begin{equation}
B_{\mu\nu}= \left(\nabla^\rho\nabla^\sigma +
    \frac{1}{2} R^{\rho\sigma}\right) C_{\mu\rho\nu\sigma}
\end{equation}
is the tracefree Bach tensor. It is the only conformally invariant tensor that is algebraically independent of the Weyl tensor.

One can write a static metric as follows
\begin{equation}
ds_4^2 = -\lambda^2\, dt^2 + h_{ij}\,dx^i dx^j,
\end{equation}
where $\lambda$ and $h_{ij}$ are functions of the spatial coordinates $x^i$.
In \cite{Lu:2015cqa} it was shown that, taking the trace of the field equations (\ref{eoms}) and integrating the equations of motion over the spatial domain from the event horizon to infinity, one can find that
\begin{equation}
\int \sqrt{h}\, d^3 x\Big[D^i(\lambda R D_i R) - \lambda (D_i R)^2 -
    m_0^2 \lambda R^2\Big]=0\,,
\end{equation}
where $D_i$ is the covariant derivative with respect to the spatial 3-metric $h_{ij}$.

By definition, $\lambda$ vanishes on the event horizon, so that if $D_i R$ goes to zero sufficiently rapidly at spatial infinity, then the total derivative term can be discarded and any static black-hole solution of (\ref{HDGaction}) must have vanishing Ricci scalar $R=0$.
The latter means that, without loss of generality, we can be constrained by the Einstein-Weyl gravity ($\beta=0$). Then, since $B_{\mu\nu}$ is tracefree, the trace of (\ref{eoms}) implies the vanishing Ricci scalar ($R=0$). Therefore, the Schwarzschild solution is also a solution for the Einstein-Weyl gravity.

Summarizing, when considering static solutions in the most general Einstein gravity with quadratic in curvature corrections given by (\ref{HDGaction}), one can take $\gamma =1$ and $\beta =0$ without loss of generality.

\smallskip

\section{Black Holes in Higher-Derivative Gravity}\label{sec:metric}

Here, first, we shall find a general analytical form of the non-Schwarzschild metric and then expand it in terms of the small deviation $k$ from the Schwarzschild branch.

\subsection{Analytical approximation}

The line element of a black hole is given by
\begin{equation}
\label{metric}
ds^2=-h(r)dt^2+\frac{dr^2}{f(r)}+r^2(d\theta^2+\sin^2\theta d\phi^2)\,,
\end{equation}
where the functions $h(r)$ and $f(r)$ satisfy the Einstein-Weyl equations of motion, which have the following form:
\begin{widetext}
\begin{eqnarray}\label{heq}
h''(r)&=&\frac{4 h(r)^2 \left(1-r f'(r)-f(r)\right)-r h(r)\left(rf'(r)+4f(r)\right) h'(r)+r^2 f(r) h'(r)^2}{2 r^2 f(r) h(r)}\,,\\
\label{feq}
f''(r)&=&\frac{h(r)-f(r) h(r)-r f(r) h'(r)}{\alpha f(r) \left(2 h(r)-r h'(r)\right)}-\frac{h(r) \left(3 r^2 f'(r)^2+12 r f(r) f'(r)-4 r f'(r)+12 f(r)^2-8 f(r)\right)}{2 r^2 f(r) \left(r h'(r)-2 h(r)\right)}\\\nonumber&&+\frac{f(r) \left(r^2 h'(r)^2-r h(r) h'(r)-2 h(r)^2\right)-r h(r) f'(r) \left(r h'(r)+4 h(r)\right)}{2 r^2 h(r)^2}\,.
\end{eqnarray}
\end{widetext}
The metric functions $f(r)$ and $h(r)$ can be expanded into the Taylor series near the horizon $r_0$:
\begin{equation}\label{horexpand}
\begin{array}{rcl}
h(r) &=& c\left[(r-r_0) + h_2(r-r_0)^2 + \ldots\right]\\
f(r) &=& f_1(r-r_0) + f_2(r-r_0)^2 + \ldots,
\end{array}
\end{equation}
where $f_1$ is the shooting parameter, which we choose in such a way that the solution is asymptotically flat, and
$c$ is the arbitrary scaling factor, which we choose such that $t$ is the time coordinate of a remote observer, i.e.,
$$\lim_{r\to\infty}h(r)=1.$$

Substituting (\ref{horexpand}) into (\ref{heq}) and (\ref{feq}), one can express all the coefficients in terms of $c$ and $f_1$, which we find using numerical integration as prescribed in \cite{Lu:2015cqa}.

It is useful to introduce the dimensionless parameter, which parameterizes the solutions up to the rescaling
\begin{equation}
p=\frac{r_0}{\sqrt{2\alpha}}.
\end{equation}
Notice that for all $p$ the Schwarzschild metric is the exact solution of the Einstein-Weyl equations as well, but at some minimal nonzero $p_{min}$, in addition to the Schwarzschild solution, there appears the non-Schwarzschild branch (found numerically in \cite{Lu:2015cqa}) which describes the asymptotically flat black hole, whose mass is decreasing, when $p$ grows, and vanishing at some $p_{max}$. The approximate maximal and minimal values of $p$ are:
\begin{equation}
p_{min} \approx 1054/1203 \approx 0.876, \quad p_{max} \approx 1.14
\end{equation}

Following the parametrization procedure given in \cite{Rezzolla:2014mua} we define the functions $A$ and $B$ through the following relations:
\begin{eqnarray}\label{hd}
h(r)&\equiv&xA(x)\,,\\
\label{fd}
\frac{h(r)}{f(r)}&\equiv&B(x)^2\,,
\end{eqnarray}
where $x$ denotes the dimensionless compact coordinate
\begin{equation}
x \equiv 1-\frac{r_0}{r}\,.
\end{equation}
We represent the above two functions as follows:
\begin{eqnarray}
A(x)&=&1-\epsilon (1-x)+(a_0-\epsilon)(1-x)^2+{\tilde A}(x)(1-x)^3,\nonumber\\
\label{asympfix}
B(x)&=&1+b_0(1-x)+{\tilde B}(x)(1-x)^2,
\end{eqnarray}
where ${\tilde A}(x)$ and ${\tilde B}(x)$ are introduced in terms of the continued fractions, in order to describe the metric near the event horizon $x=0$:
\begin{align}\nonumber
{\tilde A}(x)=\frac{a_1}{\displaystyle 1+\frac{\displaystyle
    a_2x}{\displaystyle 1+\frac{\displaystyle a_3x}{\displaystyle
      1+\frac{\displaystyle a_4x}{\displaystyle
      1+\ldots}}}}\,,\\\label{contfrac}
{\tilde B}(x)=\frac{b_1}{\displaystyle 1+\frac{\displaystyle
    b_2x}{\displaystyle 1+\frac{\displaystyle b_3x}{\displaystyle
      1+\frac{\displaystyle b_4x}{\displaystyle
      1+\ldots}}}}\,.
\end{align}
At the event horizon one has: ${\tilde A}(0) = a_1, \quad {\tilde B}(0) =b_1$.


We notice that (\ref{heq}) and (\ref{feq}) imply that $a_0=b_0=0$, i.e. the post-Newtonian parameters for the non-Schwarzschild solution coincide with those in General Relativity.
We fix the asymptotic parameter $\epsilon$ as
\begin{equation}
\epsilon=-\left(1-\frac{2M}{r_0}\right)\,,
\end{equation}
using the value of the asymptotic mass which can be found by numerical fitting of the asymptotical behavior of the metric functions.

Expanding (\ref{hd}) and (\ref{fd}) near the event horizon we find the parameters $a_1,a_2,a_3,\ldots$,$b_1,b_2,b_3,\ldots$ as functions of $c$ and $f_1$. In their turn, the values of $c$ and $f_1$ can be found numerically for each value of the parameter $p$. It appears that $\epsilon$, $a_1$, and $b_1$ approach zero for
$$p\approx\frac{1054}{1203},$$
where the numerical non-Schwarzschild solution coincides with the Schwarzschild one.

The fitting of numerical data for various values of $p$ and $r_0$ shows that $\epsilon$, $a_1$, and $b_1$ can be approximated within the maximal error $\lesssim 0.1\%$ by parabolas as follows:
\begin{eqnarray}
\label{ep} \epsilon&\approx&(1054 - 1203 p)\left(\frac{3}{1271} + \frac{p}{1529}\right)\,,\\
\label{a1p} a_1&\approx&(1054 - 1203 p)\left(\frac{7}{1746}-\frac{5 p}{2421}\right)\,,\\
\label{b1p} b_1&\approx&(1054 - 1203 p)\left(\frac{p}{1465}-\frac{2}{1585}\right)\,.
\end{eqnarray}

These fittings are almost linear in $p$.
Indeed, if one uses $k = 1054-1203 p$, then the above relations read
\begin{eqnarray}
 \epsilon&\approx&\frac{6857795 }{2337860877}\, k \left(1-\frac{1271}{6857795}\, k\right)+\Order{k^3}\,,\\
 a_1&\approx&\frac{1242869 }{565017822}\, k \left(1+\frac{970}{1242869 } \, k\right) +\Order{k^3}\,,\\
 b_1&\approx&-\frac{370840 }{558679215} \, k \left(1+\frac{317}{370840} \, k\right)+\Order{k^3}\,,
\end{eqnarray}
where one can see that the coefficients of the quadratic form are quite small. Nevertheless they cannot be neglected if one aims at $0.1\%$ accuracy.

\begin{figure}
\resizebox{\linewidth}{!}{\includegraphics*{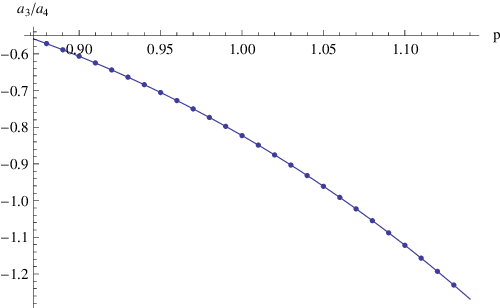}}
\caption{Fit of the numerically found parameters. The ratio $a_3/a_4$ and the corresponding fit, which remains finite for any $p\geq\frac{1054}{1203}$.}\label{fig:fit}
\end{figure}
\begin{figure*}
\resizebox{\linewidth}{!}{\includegraphics*{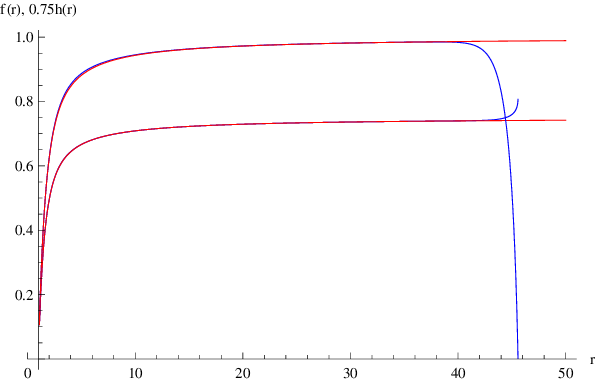}\includegraphics*{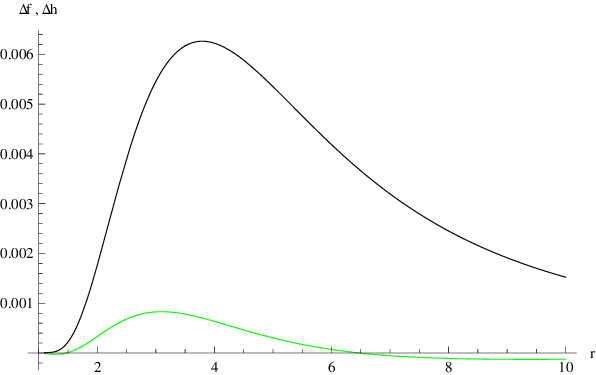}}
\caption{Comparison of numerical and analytical approximations for the metric functions: $r_0=1$, $\alpha=0.5$ ($p=1$). Left panel: $f(r)$ (upper) and, rescaled, $h(r)$ (lower). Numerical approximation (blue) fails at sufficiently large distance while our analytical approximation (red) has the correct behavior both near and far from the event horizon. Right panel: the difference between analytical and numerical approximations for $f(r)$ (black, upper) and $h(r)$ (green, lower). The largest difference is around the innermost stable circular orbit of a massive particle and photon circular orbit, where it still remains smaller than $0.1\%$.}\label{fig:metric}
\end{figure*}

With the same accuracy we are able to find $a_2$ and $b_2$ as
\begin{eqnarray}
\label{a2p} a_2&\approx&\frac{6 p^2}{17}+\frac{5 p}{6}-\frac{131}{102}\,,\\
\label{b2p} b_2&\approx&\frac{81 p^2}{242}-\frac{109 p}{118}-\frac{16}{89}\,.
\end{eqnarray}
$a_3$ and $a_4$ diverge at
$$p\approx\frac{237}{223}.$$
Therefore we find out that these parameters can be well approximated as
\begin{eqnarray}
\label{a3p} a_3&\approx&\frac{\dfrac{9921 p^2}{31}-385 p+\dfrac{4857}{29}}{237-223 p}\,,\\
\label{a4p} a_4&\approx&\frac{\dfrac{9 p^2}{14}+\dfrac{3149 p}{42}-\dfrac{2803}{14}}{237-223 p}\,.
\end{eqnarray}
In this way, although each of the parameters $a_3$ and $a_4$ diverges at $p \approx 237/223$, its ratio is finite and thereby has finite contribution into the continued fraction (see fig.~\ref{fig:fit}).
Finally, we observe that $b_3$ and $b_4$ are well approximated by the straight lines as
\begin{eqnarray}
\label{b3p} b_3&\approx&-\frac{2 p}{57}+\frac{29}{56}\,,\\
\label{b4p} b_4&\approx&\frac{13 p}{95}-\frac{121}{98}\,.
\end{eqnarray}

Within the chosen accuracy of a fraction of $0.1\%$ for the metric functions $f(r)$ and $h(r)$ (see fig.~\ref{fig:metric}) we can set $a_5=b_5=0$ in (\ref{contfrac}) and, substituting the found above coefficients $\epsilon$, $a_1$, $a_2$, $a_3$, $a_4$, $b_1$, $b_2$, $b_3$, $b_4$ into (\ref{asympfix}), obtain the final analytic expressions for the metric functions as the forth order continued fraction expansion (see appendix~\ref{sec:metricform}).

If one is limited by a rougher approximation of the second order, all the parameters $\epsilon$, $a_1$, $b_1$, $a_2$, $b_2$ can also be well approximated by linear (in $p$) polynomials instead of the quadratic ones. Such an approximation would have much simpler analytical form (which will be discussed in subsection C), leading to the larger maximal error of about a few percents.

\smallskip

\subsection{Expansion of the forth order approximation near the Schwarzschild solution}
When one is interested in relatively small deviations from the Schwarzschild geometry, a more concise expressions can be obtained by using the expansion in terms of $k$. In order to have a positive asymptotic mass, the values of $k$ can vary from $0$ until approximately $-321.727$. For example, the final formula for $h(r)$ expanded up to the first order in $k$ then reads
\begin{equation}\label{hkexp}
h(r) =\left(1-\frac{r_{0}}{r}\right)\left(1 - k\frac{r_{0}}{r} \frac{h_{1}(r)}{h_{2}(r)}+\Order{k^2}\right),
\end{equation}
where
\begin{widetext}
\begin{eqnarray}
h_{1}(r) &=& 7094296364854698294656777815 r^3+ 2700140790021572890363934045 r^2 r_{0}
\\\nonumber&&+32852984866789222219083981378 r r_{0}^2-4194480693404458083513273360 r_{0}^3,\\
h_{2}(r) &=& 61001803863561 r \left(39646131244569649 r^2-24556525364789942 r r_{0}+156809140779977329 r_{0}^2\right).
\end{eqnarray}
This formula is considerably shorter than the full formula for $h(r)$, what might be useful when one numerically models various process around the black hole. The ratio of both metric functions is
\begin{equation}
\frac{f(r)}{h(r)} = 1+\frac{37793605455056 k r_{0}^2 (50038777 r+84360383 r_{0})}{278643 r \left(6266529735540821295 r^2-3742896005107026923 r
   r_{0}+11207698915983181988 r_{0}^2\right)}+ \Order{k^2}.
\end{equation}
\end{widetext}
When $|k| \lessapprox 100$, first order expansions in $k$ for $h(r)$ and $f(r)$ stay within a few tenths of one percent from the full analytical metric, keeping thereby the same order of the general error. The $k$-expanded metric is also included into the Mathematica\textregistered{} notebook we share.

\subsection{Second order approximation: more compact, but robust analytical metric}
In case one is interested in a much more robust, but compact expression for the metric, one can be limited by the second order in the expansion~(\ref{contfrac}), i.~e. take $a_3=b_3=0$. Then, it is sufficient to consider a linear fit for $\epsilon$, $a_1$, $a_2$, $b_1$, $b_2$ as follows
\begin{eqnarray}
\label{ep2} \epsilon&\approx&\frac{1054 - 1203 p}{326}\,,\\
\label{a1p2} a_1&\approx&\frac{1054 - 1203 p}{556}\,,\\
\label{a2p2} a_2&\approx&-\frac{18 - 17 p}{11}\,,\\
\label{b1p2} b_1&\approx&-\frac{1054 - 1203 p}{1881}\,,\\
\label{b2p2} b_2&\approx&-\frac{2+p}{4}\,.
\end{eqnarray}

Thus, we obtain even simpler (than in the two previous subsections) form for the metric functions $A(r)$ and $B(r)$ in (\ref{appmetr}),
\begin{eqnarray}
A(r)&=&1-\frac{(1054 - 1203 p)r_0^2}{2r^2}\times
\\\nonumber&&\times\left(\frac{r+r_0}{163r_0}+\frac{11r_0}{278(7r-18r_0-17p(r-r_0))}\right)\,,\\
B(r)&=&1-\frac{4(1054 - 1203 p)r_0^3}{1881r^2(2(r+r_0)-p(r-r_0))}\,.
\end{eqnarray}
Yet, the obtained metric is considerably less accurate: for $p< 0.97$ the relative error stays within a fraction of one percent, but for near extremal values of $p$ it may reach a few percents. The accuracy of the approximation for a given value of $p$ of the second and forth order approximation can be learnt from the Mathematica notebook we share with readers.

\section{Testing the accuracy of the approximation}\label{sec:accuracy}

The metric is not a gauge-invariant characteristic and, strictly speaking, comparing the metric functions has no direct physical interpretation. Therefore, the best way to test the accuracy of the analytical metric obtained in the previous section is to calculate basic observable quantities for the analytical metric and compare them with the accurate ones found for the numerical metric. Here we shall consider two kinds of such observable characteristics: the frequency of a massive particle on the innermost stable circular orbit (ISCO) and frequencies of the quasinormal modes in the eikonal (short wavelength) regime.

\subsection{Innermost stable circular orbit}

First, we shall compute the radius of the smallest circular orbit of a massive test particle rotating around the black hole.
The circular movement of a massive particle is described by the following potential
$$V_m(r)=\frac{E^2}{h(r)}-\frac{L^2}{r^2}-1,$$
where $E$ and $L$ are, respectively, the energy and momentum per unit mass.

The innermost stable circular orbit corresponds to
$$V_m(r)=V_m'(r)=V_m''(r)=0,$$
which is reduced to the following equation for the radial coordinate of the orbit
$$r h(r) h''(r)-2 r h'(r)^2+3 h(r) h'(r)=0.$$
We solve the above equation numerically with $h(r)$ given in the Appendix A in the analytical from and in \cite{Lu:2015cqa} numerically.

The corresponding orbital frequencies are given
\begin{equation}\label{Omega}
\Omega=\sqrt{\frac{h'(r)}{2r}}.
\end{equation}

\begin{figure}
\resizebox{\linewidth}{!}{\includegraphics*{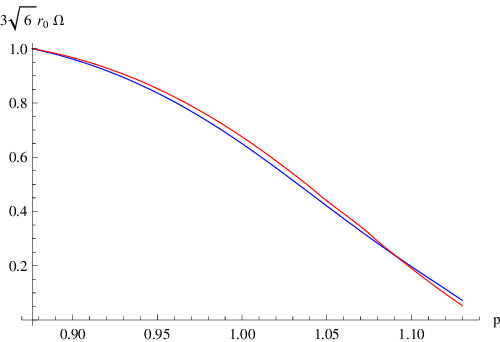}}
\caption{ISCO frequency $\Omega$ normalized by the Schwarzschild value: red (upper) is for analytical metric and blue (lower) is for the numerical one.}\label{fig:ISCO}
\end{figure}
From Fig.~\ref{fig:ISCO} we see that 
the frequency of ISCO decreases quite a few times, as the parameter $p$ grows. This means that the ISCO moves outward the black hole at a great extent. The relative error stays within a few percents for the parametric region under consideration, being much less for the near Schwarzschild and near extremal (almost massless) cases. At the same time, for the intermediate values of $p$, when the error of the analytical approximation is maximal (see Fig.~\ref{fig:ISCO}), the effect of the deviation from the Schwarzschild metric on the orbital frequency is already much larger than the error. This means that the forth order approximation developed here is adequate.

\subsection{Analytical formulas for the eikonal quasinormal frequencies}

Here we shall consider the proper oscillations frequencies, called \emph{quasinormal modes} \cite{Konoplya:2011qq}, of a test field in the background of the black hole in the high frequency (eikonal or high multipole number) regime. The boundary conditions for the quasinormal modes are purely incoming wave on the event horizon and purely outgoing wave at infinity. In the geometrical optic (eikonal) regime perturbations of \emph{test} fields of any spin are dominated by the same centrifugal-like part of the effective potential. Therefore, it is sufficient to consider here the derivations only for the test massless scalar field, while the resultant eikonal formulas for the test fields of other spin will be the same.
Perturbations of a test scalar field obey the general relativistic Klein-Gordon equation
\begin{equation}
\frac{1}{\sqrt{-g}}\partial _{\mu }\left( \sqrt{-g}g^{\mu \nu }\partial
_{\nu } \Phi \right) = 0 \,, \label{KGNM}
\end{equation}%
Implying that
$$\Phi(t,r,\theta,\phi)= e^{-\imo\omega t}Y_{\ell}(\theta,\phi)\Psi(r)/r,$$
where $Y_{\ell}(\theta,\phi)$ are spherical harmonics, the Klein-Gordon equation can be reduced to the following form:

\begin{figure*}
\resizebox{\linewidth}{!}{\includegraphics*{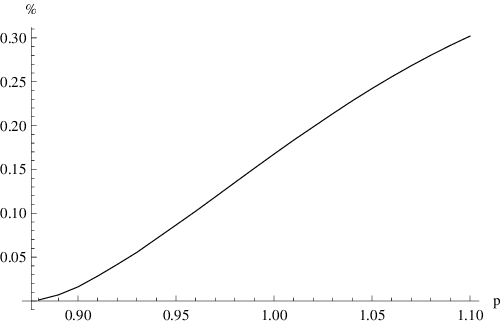}\includegraphics*{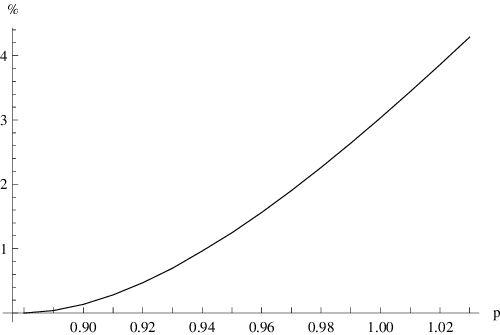}}
\caption{Relative error (in percents) of the first-order in $k=1054-1203 p$ approximated eikonal formula (\ref{eikonal}) for the real (left panel) and imaginary (right panel) parts of the eikonal formula for the QNMs.}\label{fig:error}
\end{figure*}
\begin{equation}\label{wavelike}
\frac{d^2\Psi}{dr_*^2}+\left(\omega^2-V(r_*)\right)\Psi=0.
\end{equation}
Here, $\omega$ is the frequency; the tortoise coordinate $r_*$ is defined as follows
$$dr_*=\frac{dr}{\sqrt{f(r)h(r)}},$$
and the effective potential for the large value of the multipole number $\ell$ takes the form
\begin{equation}\label{effpot}
V(r)=\left(\ell+\frac{1}{2}\right)^2\left(\frac{h(r)}{r^2}+\Order{\frac{1}{\ell^2}}\right)\,.
\end{equation}

From (\ref{hkexp}) we find that the effective potential (\ref{effpot}) has the maximum at
$$r_m=\frac{3r_0}{2}\left(1+0.000393k\right)+\Order{k^2,\ell^{-2}}.$$
As $k$ is negative, then the peak of the effective potential is closer to the black hole horizon for non-Schwarzschild solution than for the Schwarzschild one.

At high $\ell$, once the effective potential has the form of the potential barrier, falling off at the event horizon and spacial infinity, the WKB formula found in \cite{Schutz:1985zz} (for improvements and extensions of this formula, see \cite{Iyer:1986np,Kokkotas:1991,Konoplya:2003ii,Matyjasek:2017psv}) can be applied for finding quasinormal modes:
\begin{equation}
\omega^2=V(r_m)-\imo\left(n+\frac{1}{2}\right)\sqrt{-2\frac{d^2V(r_m)}{dr_*^2}}, \label{wkb}
\end{equation}
which depends on the value and the second derivative of the potential in its maximum $r_{m}$.
Using the above eikonal formula for the QNMs we find that
\begin{eqnarray}\label{eikonal}
\omega&=&\frac{2}{3\sqrt{3}r_0}\Biggr[\left(\ell+\frac{1}{2}\right)\left(1-0.001308k\right)
\\\nonumber&&-\imo\left(n+\frac{1}{2}\right)\left(1-0.002743k\right)\Biggr]+\Order{k^2,\ell^{-1}},
\end{eqnarray}
implying higher real (photon circular orbit) frequency and faster decay of the oscillations for the non-Schwarzschild branch. When $k=0$ the above formula is reduced to its Schwarzschild analogue.

We compare the above formulas with the precise values for the real and imaginary parts of the frequencies, obtained by substituting the numerical solutions into the eikonal WKB formula. From Fig.~\ref{fig:error} we see that the first-order in $k$ formula (\ref{eikonal}) provides quite an accurate result for the real (fractions of a percent) and imaginary (a few percents) parts of the eikonal frequencies.
Notice, that the obtained analytical form in terms of the deviation from the Schwarzschild solution $k$, allowed us to find easily the concise analytical non-Schwarzschild generalization of the well-known eikonal formulas for quasinormal modes of the Schwarzschild black hole.
Usually the obtained here eikonal quasinormal modes of a test scalar field determine the parameters of the circular null geodesic: the real and imaginary parts of the quasinormal mode are multiples of the frequency and instability timescale of the circular null geodesics respectively.
However, quasinormal modes of non-test, e. g., gravitational, fields may not obey this rule \cite{Konoplya:2017wot}.

\section{Discussion}\label{sec:discussion}

In the present paper we have obtained the approximate analytic expression for the black-hole solution of the non-Schwarzschild metric in the most general Einstein gravity with quadratic in curvature corrections. The obtained analytical metric represents asymptotically flat black hole which has the same post-Newtonian behavior as in General Relativity, but is essentially different in the strong field regime.
The metric is expressed in terms of event horizon radius $r_0$ and the dimensionless parameter $p = r_{0}/\sqrt{2 \alpha}$, where $\alpha$ is the coupling constant. The minimal value of $p\approx0.876$ corresponds to the merger of the Schwarzschild and non-Schwarzschild solutions, while at $p\approx1.14$ the black-hole mass approaches zero ($\epsilon=-1$).

As the metric is written in terms of the black-hole parameter and coupling constant and is accurate \emph{everywhere} outside the event horizon, it can be used in the study of the basic properties of the black hole and the description of various phenomena in its vicinity in the same way as the exact analytical solution. Our main future aim is to generalize the obtained analytical metric to the case of rotating black holes \cite{inprogress}. At the same time a number of other appealing problems associated with the obtained metric could be solved:

\begin{itemize}
\item Perturbations and analysis of stability of the non-Schwarzschild black hole;
\item Quasinormal modes of gravitational and test fields in its vicinity (this was partially done for the numerical solution in \cite{Cai:2015fia} and comparison between analytical and numerical metrics would be appealing). As higher curvature corrections frequently lead to a new branch of non-perturbative (in coupling constant) modes \cite{highercurvatureQNM}, it is interesting to check whether this phenomena takes place for the considered here quadratic gravity.
\item Analysis of massless and massive particles' motion, binding energy, innermost stable circular orbits, stability of orbits;
\item Analysis of the accretion disks and the corresponding radiation in the electromagnetic spectra;
\item Consideration of tidal and external magnetic fields in the vicinity of a black hole, etc.
\item Hawking radiation in the semiclassical and beyond semiclassical regimes;
\item Detailed study of the black-hole thermodynamics.
\end{itemize}

The obtained here analytical approximation for the metric (given in Appendix A) has two evident advantages over the numerical solution. First, it allows one to solve all the above enumerated problems in a much more economic and elegant way. Second, the analytical metric allows applications of a greater variety of methods for its analysis. For example, in order to get the full knowledge of the characteristic quasinormal spectrum of a black hole, one has to use the Leaver method \cite{Leaver} which simply cannot be applied to the numerical interpolation function and \emph{requires} the analytically written metric. Applications of other methods, for example (used here for illustration) WKB method \cite{Schutz:1985zz,Iyer:1986np} or the time-domain integration \cite{timedomain}, are considerably constrained and give information only about the lowest modes.

In addition, we found the eikonal quasinormal frequencies of test fields and the frequency and positions of ISCO for the non-Schwarzschild black hole solution in the higher derivative gravity. Comparison of the data obtained for the analytical and numerical metrics allowed us to test the accuracy of our approximation. It is shown that the non-Schwarzschild black hole is characterized by a much further position of ISCO and much slower rotational frequency of a massive particle. The eikonal quasinormal modes of the non-Schwarzschild black hole have smaller real oscillation frequencies and damping rates.

Here we used the expansion up to the forth order and achieved accuracy in the metric functions with the maximal error of fractions of a percent. Once it is necessary, expansion to higher orders will produce much more accurate representation of the metric.

\acknowledgments{R. K. would like to thank H. Lü for sharing his Mathematica\textregistered{} code which produces the numerical solution considered here.
R. K. was supported by ``Project for fostering collaboration in science, research and education'' funded by the Moravian-Silesian Region, Czech Republic and by the Research Centre for Theoretical Physics and Astrophysics, Faculty of Philosophy and Science of Silesian University at Opava. A.~Z. thanks Conselho Nacional de Desenvolvimento Científico e Tecnológico (CNPq) for support and Theoretical Astrophysics of Eberhard Karls University of Tübingen for hospitality.}

\appendix

\begin{widetext}
\section{Analytical form of the metric functions}\label{sec:metricform}

Here we provide the obtained analytical form of the metric. In the attachment to this article we share with readers the Mathematica notebook, where the metric functions are explicitly written down.

The black-hole metric can be written in the form:
\begin{equation}\label{appmetr}
ds^2=-\left(1-\frac{r_0}{r}\right)A(r)dt^2+\frac{B(r)^2dr^2}{\left(1-\dfrac{r_0}{r}\right)A(r)}+r^2(d\theta^2+\sin^2\theta d\phi^2)\,,
\end{equation}

Here, we present the metric functions $A(r)$ and $B(r)$ in terms of the parameters $b$ and $r_0$. Notice, that one can get the black hole radius $r_0 =1$ (which leads to the re-definition of the black-hole mass) and express everything in terms of $r_0$.
\begin{eqnarray}\nonumber
A(r)&=&\Biggr[152124199161 \left(873828 p^4-199143783 p^3+806771764 p^2-1202612078 p+604749333\right) r^4
\\\nonumber&&+78279\left(1336094371764p^6-300842119184823 p^5+393815823540843 p^4+2680050514097926 p^3\right.
\\\nonumber&&\left.-9501392159249689 p^2+10978748485369369 p-4249747766121792\right)r^3 r_0
\\\nonumber&&-70372821 \left(1486200636 p^6+180905642811 p^5+417682197141 p^4-1208134566031 p^3\right.
\\\nonumber&&\left.-324990706209 p^2+3382539200269p-2557857695019\right) r^2 r_0^2-\left(104588131327314156 p^6\right.
\\\nonumber&&-23549620247668759617 p^5-435688050031083222417p^4+2389090517292988952355 p^3
\\\nonumber&&\left.-3731827099716921879958 p^2+2186684376605688462974 p-389142952738481370396\right) r r_0^3
\\\nonumber&&+31\left(3373810687977876 p^6+410672271594465801 p^5-14105000476530678231 p^4+51431640078486304191 p^3\right.
\\
&&\left.-71532183052581307042p^2+43250367615320791700 p-9476049523901501640\right) r_0^4\Biggr]
\\\nonumber&&/\Biggr[152124199161 r^2\Biggr(\left(873828 p^4-199143783 p^3+806771764 p^2-1202612078 p+604749333\right) r^2
\\\nonumber&&-2 \left(873828 p^4-47583171 p^3+386036980 p^2-678598463 p+341153481\right) r r_0
\\\nonumber&&+899\left(972 p^4+115659 p^3-38596 p^2-1127284 p+1101579\right) r_0^2\Biggr)\Biggr]\,,
\end{eqnarray}
\begin{eqnarray}\nonumber
B(r)&=&\Biggr[464405 \left(3251230164 p^3-14548777134 p^2+20865434326 p+23094914865\right) r^3
\\\nonumber&&-464405 \left(6502460328 p^3-52856543928p^2+100077612184 p-32132674695\right) r^2 r_0
\\\nonumber&&-\left(1244571650887908 p^3+17950319416564777 p^2-53210739821255918p+5097428297648940\right) r r_0^2
\\
&&+635371 \left(4335198168 p^3-42352710803 p^2+90235778452 p-49464019740\right)r_0^3\Biggr]
\\\nonumber&&/\Biggr[464405 r \Biggr(\left(3251230164 p^3-14548777134 p^2+20865434326 p+23094914865\right) r^2
\\\nonumber&&-\left(6502460328p^3-52856543928 p^2+100077612184 p-32132674695\right) r r_0
\\\nonumber&&+6\left(541871694 p^3-6384627799 p^2+13202029643p+2626009760\right) r_0^2\Biggr)\Biggr]\,.
\end{eqnarray}
\end{widetext}

Here, the minimal value of $p \approx 1054/1203\approx0.876$ corresponds to the merger of the Schwarzschild and non-Schwarzschild solutions and at the maximal value of $p\approx1.14$ the black-hole mass approaches zero ($\epsilon=-1$).

\end{document}